%% file: ms.tex
\documentclass[prx,superscriptaddress,floatfix,nopacs,nofootinbib]{revtex4-2}
\usepackage{graphicx,amsfonts,amssymb,amsmath,hyperref,enumerate}

\usepackage[ruled,lined]{algorithm2e}
\usepackage{pgfplots}
\usepackage{pgf}
\usepackage{lmodern}
\usepackage{import}
\usepackage{xr}
\usepackage{enumitem}
\usepackage{soul}

\usepackage{amsthm}

\usepackage{cleveref}
\usepackage{autonum}

\usepackage{bbm}
\usepackage{dsfont}
\usepackage{diagbox}

\usepackage{tikz}
\usepackage{subfigure}

\newif\ifhyper
\ifhyper
\hypersetup{
   citecolor = {green},
   colorlinks = {true}, 
   urlcolor = {blue} 
}
\fi

\newcommand{\beq}{\begin{equation}}
\newcommand{\eeq}{\end{equation}}
\newcommand{\beqa}{\begin{eqnarray}}
\newcommand{\eeqa}{\end{eqnarray}}

\newcommand{\comment}[1]{}

\newcommand{\hrho}{\hat{\rho}}
\newcommand{\htau}{\hat{\tau}}

\newcommand{\cV}{\mathcal{V}}
\newcommand{\cC}{\mathcal{C}}

\newcommand{\cF}{\mathcal{F}}
\newcommand{\cB}{\mathcal{B}}
\newcommand{\cX}{\mathcal{X}}

\def\Longarrow{\protect\@lra}
\def\@lra{\relbar\joinrel\relbar\joinrel\relbar\joinrel%
          \relbar\joinrel\rightarrow}

\newtheorem{definition}{Definition}

\newtheorem{thm}{Theorem}

\newtheorem{rem}{Remark}

\begin{document} 

\title{Modeling Stochastic Data Using Copulas For Application in Validation of Autonomous Driving}

\author{Katrin Lotto}
\affiliation{ZF Friedrichshafen AG,
	Research and Development, Graf-von-Soden-Platz 1,
	88046 Friedrichshafen,
	Germany.}
\author{Thomas Nagler}
\affiliation{Department of Statistics, LMU Munich, 
	80799 Munich,
	Germany}
\author{Mladjan Radic}
\affiliation{ZF Friedrichshafen AG,
Research and Development, Graf-von-Soden-Platz 1,
88046 Friedrichshafen,
Germany.}

\begin{abstract}
\noindent Verification and validation of fully automated vehicles is linked to an almost interactable challenge of reflecting the real world with all its interactions in a virtual environment. Influential stochastic parameters need to be extracted from real-world measurements and real-time data, capturing all interdependencies, for an accurate simulation of reality. A copula is a probability model that represents a multivariate distribution, examining the dependence between the underlying variables. This model is used on drone measurement data from a roundabout containing dependent stochastic parameters. With the help of the copula model, samples are generated that reflect the real-time data. Resulting applications and possible extensions are discussed and explored.
\end{abstract}

\maketitle

\input{introduction}
\input{preliminaries}

\input{numerical_examplesRound}

\input{conclusion}

\section*{Acknowledgement}
The research leading to these results is funded by the German Federal Ministry for Economic Affairs and Climate Action within the project ``Verifikations- und Validierungsmethoden automatisierter Fahrzeuge im urbanen Umfeld''. The authors would like to thank the consortium for the successful cooperation

\bibliography{bibliography.bib}
\end{document}

%% file: introduction.tex
\section{Introduction}

Autonomous driving vehicles have a huge potential not only for reducing the number of road accidents but also from an economic point of view, reducing emissions as well as increasing the efficiency of traffic usage by, e.g., robotaxis or avoiding traffic jams. Thus autonomous driving already revolutionizes our view on mobility, traffic safety and how time is spent during car journeys. The question arises, how to verify that an autonomous vehicle is safe enough. There are many proposals of how many kilometers an autonomous vehicle has to drive accident-free before it may be safely released on the road. The estimation reaches from several millions to even billions of kilometers. We want to stress, that this is only for one system. Every update or change of the software, occuring errors, improvements and transformations require a rerun. Every experiment of such dimension, if feasible at all, would be tremendously cost-intensive. This motivates a simulation based testing of autonomous vehicles. 

A simulation based framework aims to replicate the real-world traffic and to validate the automated driving functionality accordingly. In this framework, the robustness, reliability and safety can be investigated in a much quicker and cost-efficient way. It is emphasized, that the validation results and investigations in the simulated environment are only as trustworthy as the virtual environment reflects the real-world \cite{behrensdorf, andrade, tang, zhao}.

The simulation is fed with stochastic data, e.g. velocity/acceleration of the traffic participants, extracted from real recordings of traffic events.

A huge challenge is that the stochastic parameters, such as velocities of every traffic participants are not independent. There are a lot of factors, which have to be considered. The weather,  visibility, amount of traffic participants are correlated to the driving behaviour of the participants. Unfortunately, this correlation is unidentified and cannot be assumed beforehand. In this work, dependencies in stochastic data are investigated and described by so called copulas.

The latin word \emph{copulare} may be translated in ``to join'' or ``to connect'', which already reveals the meaning and functionality of copulas. They are a great tool for modeling the (nonlinear) dependence of random variables and to calculate the joint distribution. The first appearance of the term copula may be found in \cite{sklar}. Since then the wide research and application of copulas can be seen in the fields of statistics \cite{czado,joe,sklar}, economics \cite{LI2021103865,kielmann}, finance \cite{brechmann,rank2007copulas,cherubini2004copula,cherubini2011dynamic}, actuarial science \cite{copulasRiskVersicherungen,gschlossl2007spatial} or risk management \cite{goodwin2015copula,ZHANG2019376,karimalis2018measuring}, just to name a few.

The outline of this work is as follows. In Section 2, we will go into further depth of how to validate automated vehicles and the theory of copulas. The construction and usage for resimulating data is described. Section 3 illustrates how to extract data from observations and how to apply the theory on a concrete numerical example and application. A conclusion and outlook for further research and development is given in Section 4.

%% file: preliminaries.tex
\section{Preliminaries}

\subsection{Safety validation of automated vehicles}

Unlike today's cars, autonomous vehicles require additional regulations. A distinction is made between automation levels, as in the SAE J3016 standard \cite{SAE}: a \textit{Driver-Only-Vehicle} without automation, \textit{assisting systems} and \textit{semi-automated systems}. In all three levels the driver keeps the responsibility for the behavior of the vehicle. Furthermore, we speak of \textit{fully automated vehicles} when the driver no longer acts as a fallback \cite{winner}. In this case, the automated systems should take over vehicle control permanently and should act safely in any environment. 

The international organisation for standardisation (ISO) provides the ISO/PAS(Publicly Available Specification) 21448 to ensure the safety of the functionalities \cite{ISOPAS} and the ISO 26262 to ensure, that no hazards are caused by technical failure \cite{ISO26262}. ISO/PAS 21448, safety of the intended funcionality (SOTIF), focuses on predictable misuse by the driver, as well as accidents that are explicitly not caused by component failure, but situations that were not planned for during development. Nevertheless, this standard does not provide a detailed strategy for identifying functional deficiencies.
In contrast, ISO 26262 focuses on safety in terms of intrinsic safety (protection of the environment from the product), therefore functional safety. It is an ISO standard for safety-related electrical/electronic systems in motor vehicles. It is not enough that the function has been executed correctly, but it must also be ensured that the function has been executed in the correct context. For example, an airbag must not be triggered when driving too fast over speed bumps. However, the minimum requirements for safeguarding an autonomous vehicle are not adequately described by this standard. The goal is to minimize risks to a level that is ``acceptable to society''. 

Consequently validation of autonomous vehicles cannot be based on existing ISO standards and need extension. Various approaches have been developed starting with Advanced Driver Assistant Systems (ADAS), where function-based approaches and real-world testing operate well. In the function-based approach, requirements are defined for the operating system that are tested by simulation or on the test track. In real-world testing, a mileage-based evaluation of functionality from field tests with a driver is performed. For both methods, validating full autonomous systems is economical infeasible in its current state. Furthermore there is the shadow mode, presented by Wang and Winner \cite{wang}, in which the automated driving function is excecuted passively in series production vehicles. The driving function receives the (real) information from the sensors, but will not act. Its actions are evaluated afterwards by simulation. 

Considering the real world as an open parameter space, with an infinite number of traffic events, the scenario-based approach tries to identify these traffic events and describe them in scenarios. It also attempts to exclude non-relevant traffic events, where neither actions nor events are observed, and to cluster similar traffic events into a representative scenario \cite{nalic}. However, this leads to the question of how to find the set of representative scenarios. A good overview of the problem of identifying critical scenarios is provided by Neurohr et al, Riedmaier et al and Zhang et al. \cite{neurohr, riedmaier, zhang}.

Menzel et al. \cite{menzel} distinguish three categories of scenarios: functional, logical and concrete scenarios. In the case of functional scenarios, the scenario space, in the same way as traffic events, is described on a sematic level by "a linguistic scenario annotation" \cite{menzel}. For the logical scenarios, this is described at the state space level. Entities and their relationships are described using parameter ranges in the state space and optionally specified using correlations and numerical relationships. A traffic event is finally mapped explicitly to the state space given a concrete scenario. Entities and their relationships are described using concrete values for each parameter. 
According to Zhang, the identification of critical scenarios can be done on all three levels of abstraction. This requires a clear definition of the operational design domain (ODD), a definition of the operating conditions under which an AD system is attempted to operate, and "the formulation of a functional scenario to a logical scenario". 

The German research project Pegasus\footnote{Pegasus - Project for the Establishment of Generally Accepted quality criteria, tools and methods as well as Scenarios and Situations for the release of highly-automated driving functions, see \url{https://www.pegasusprojekt.de/en/home}} followed the approach in \cite{schuldt} of modeling the environment in layers and extended it to 6 layers describing the environment of a highway. In the follow-up project VVM\footnote{VVM - Verification and Validation Methods for Level 4 and 5 Automated Vehicles, see \url{https://www.vvm-projekt.de/}}, the 6-layer model was refined, extended to the urban environment, provided with guidelines \cite{scholtes}. 

To describe the operating conditions of an AD system, the parameters of the six layers are used, such as weather, number of road users, number of lanes or speed of a cyclist. The assessment of whether a scenario is critical or not is based on the concrete values for these stochastic parameters. Some studies already consider realistic parameter distributions \cite{zhang} obtained from real-life driving databases.

In scenario-based testing, parameter distributions can be used to support the search for critical scenarios within the scenario space. In doing so, they serve to model the mutuality of the scenarios. In \cite{akagi}, a risk index based on a Gaussian Mixture Model is used to efficiently select critical traffic conditions. Thereafter, the scenarios are replicated via simulation and used, for example, to evaluate the criticality of the AD system. For the simulation of the scenarios, real trajectories, extracted from real data are used. In the same way, characteristic criteria of scenarios are taken from the real data, e.g. parameter distributions or parameter dependencies, and used to reproduce trajectories or parameters. Thus, parameter distributions are applied in estimating the failure rate of a scenario. For example, Wagner et al \cite{wagner} predict the behavior of road users based on conditional distributions obtained by analyzing naturalistic driving study called euroFOT (large-scale European Field Operational Test on Active Safety Systems) and determine the criticality for each predicted event.

\subsection{Copulas}

Copulas allow to model the (nonlinear) dependence of several random variables. With the help of copulas, the multivariate distribution can be split into their marginal distributions.
In the following, the definition and construction as well as the main features of copulas will be discussed and illustrated.

\subsubsection{Definition}

For a $d$-dimensional random vector $X = (X_1 , \ldots , X_d)$, we denote the joint cumulative distribution function by $F_{X}(x) = P(X_1 \leq x_1, \ldots, X_d \leq x_d)$ and the marginal distributions by $F_k(x_k) = P(X_i \leq x_i)$. A copula is a special type of distribution function.
\begin{definition}[Copulas]
	A $d$-dimensional copula $C\colon [0,1]^d \to [0,1]$ is a multivariate distribution function with standard uniform marginals. If the copula is absolutely continuous, the corresponding copula density is defined as 
	\begin{align}
		c (u_1 , \ldots , u_d) = \frac{\partial^d}{\partial u_1 \ldots \partial u_d} C(u_1,\ldots,u_d), \quad \forall
		(u_1,\ldots,u_d) \in [0,1]^d .
	\end{align}
\end{definition}

Sklar's theorem \cite{sklar} states that any multivariate distribution can be expressed in terms of its marginal distributions and a copula.
\begin{thm}[Sklar's Theorem, 1959]
	Let $X = (X_1 , \ldots , X_d)$ be a $d$-dimensional random vector and let the corresponding joint distribution function be denoted by $F$ as well as the marginal distribution functions by $F_i$, where $i=1,\ldots,d$. Then, the joint distribution function is given by
	\begin{align}
		F(x_1, \ldots , x_d) = C\left( F_1 ( x_1) , \ldots , F_d (x_d) \right),
	\end{align}
	where $C$ is a $d$-dimensional copula. If the distributions are absolutely continuous, then the copula $C$ is unique.
	Moreover, the corresponding density function is given by
	\begin{align}
		f(x_1 , \ldots , x_d) = c\left( F_1 ( x_1) , \ldots , F_d (x_d) \right) \prod_{i=1}^d f_i (x_i)
	\end{align}
\end{thm}

Sklar's Theorem can be understood as follows. The marginal distributions $F_1, \dots, F_d$ characterize the behavior of each of the variables $X_1, \dots, X_d$ in isolation. The copula on the other hand characterizes the dependence between them. On the one hand, we can decompose any joint distribution into marginal distributions and a copula. On the other, we can combine any copula with arbitrary marginal distributions to form a valid multivariate disribution function.

\newtheorem{example}{Example}

\begin{example}
	The copula corresponding to independent random variables $X_1, \dots, X_d$ is $C(u_1, \dots, u_d) = u_1 \cdots u_d$ and called \emph{independence copula}.
\end{example}

\subsubsection{Bivariate parametric copula models}

There are two major classes of parametric copula models: Elliptical copulas and Archimedean copulas. Two well-known examples of elliptical copulas are the multivariate Gauss copula and the multivariate Student-$t$ copula. They are constructed implicitly by inverting the formula in Sklar's theorem, i.e.,
\begin{align}
	C\left( u_1 , u_2 \right) = F \left( F_1^{-1} (u_1) , F_2^{-1} (u_2) \right),
\end{align}
and take $F$ to be a bivariate Gauss or Student-$t$ distribution. 
For further information, see e.g. \cite{czado}. 

Archimedean copulas are constructed more explicitly via generator functions.

\begin{definition}[Bivariate Archimedean Copulas]
	Let $\varphi:[0,1] \rightarrow [0,\infty]$ be a continuous, strictly monotone decreasing and convex function satisfying $\varphi(1)=0$, then $\varphi$ is the generator of the bivariate Archimedean copula
	\begin{align}
		C(u_1,u_2) := \varphi^{[-1]} \left( \varphi(u_1) + \varphi (u_2) \right),
	\end{align}
	where
	\begin{align}
		\varphi^{[-1]}: [0,\infty] \rightarrow [0,1] : \qquad
		\varphi^{[-1]} (t) = \begin{cases}
			\varphi^{-1} (t), & \quad 0 \leq t \leq \varphi(0), \\
			0, & \quad \varphi(0) < t \leq \infty
		\end{cases}
	\end{align}
	is the so called pseudo-inverse of $\varphi(\cdot)$.
\end{definition}
For a specific choice of the generator, the bivariate copulas given in Table \ref{table:arch} may be constructed. Note, that the BB1 and BB7 families are two-parametric famlies where the others are one-parametric Archimedean copulas, but \cite{czado} is suggested for further information.
\begin{table}[h!]
	\centering
	\begin{tabular}{|l|c|c|c|}
		\hline
		\textbf{Name} & \textbf{Bivariate Archimedean Copula} $C(u_1,u_2)$ & $\boldsymbol{\delta}$ & $\boldsymbol{\theta}$ \\
		\hline
		Clayton & $\left(u_1^{-\delta} + u_2^{-\delta} -1 \right)^{-1/\delta}$ 
		& $\delta \in (0,\infty)$ & $-$\\
		\hline
		Gumbel & $\exp \left( - \left[ \left(-\ln(u_1)\right)^{\delta} + \left(-\ln(u_2)\right)^{\delta} \right]^{1/\delta} \right)$ 
		& $\delta \geq 1$ & $-$\\
		\hline
		Frank & $ -\frac{1}{\delta} \ln\left( \frac{1}{1-e^{-\delta}} 
		\left[ (1-e^{-\delta}) - (1-e^{-\delta u_1}) (1-e^{-\delta u_2}) \right] 
		\right) $ 
		& $\delta \neq 0$ & $-$\\
		\hline
		Joe & $1 - \left[ (1-u_1)^{\delta}  + (1-u_2)^{\delta} 
		- (1-u_1)^{\delta}(1-u_2)^{\delta}
		\right]^{1/\delta}$ 
		& $\delta \geq 1$ & $-$ \\
		\hline
		BB1 & $\left\{1+\left[\left(u_1^{-\theta}-1\right)^\delta + \left(u_2^{-\theta} - 1\right)^\delta\right]^{1/\delta}\right\}^{-1/\theta}$ 
		&  $\delta \geq 1$ & $\theta > 0$\\
		\hline
		BB7 & $1-\left(1-\left[\left(1-\left(1-u_1\right)^\theta\right)^{-\delta} + \left(1-\left(1-u_2\right)^\theta\right)^{-\delta}-1\right]^{-1/\delta}\right)^{1/\theta}$ 
		& $\delta > 0$ & $\theta \geq 1$\\
		\hline
	\end{tabular}
	\caption{Classes of one- and two-parametric archimedean copulas.}
	\label{table:arch}
\end{table}
\begin{rem}
	Perfect dependence and independence respectively is achieved for Clayton if $\delta \rightarrow \infty$ and $\delta \rightarrow 0$ respectively and for Gumbel if $\delta \rightarrow \infty$ and $\delta = 1$ respectively. For the Frank copula, the indepence copula is achieved for $\delta \rightarrow 0^+$ and for the Joe copula, the indepence copula is achieved for $\delta = 1$. For the BB1 copula, the independence copula corresponds to $\theta \rightarrow 0^+$ and $\delta \rightarrow 1^+$ and for the BB7 copula, the independence copula corresponds to  $\theta = 1$ and $\delta = 0$. 
\end{rem}

The copula $C$ is related to the distribution of the random variables $U_j = F_j(X_j)$, $j = 1, \dots, d$. We thus first estimate the marginal distributions $F_1, \dots, F_d$ by $\widehat F_1, \dots, \widehat F_d$. Then we generate ``pseudo-observations'' $\widehat U_j = \widehat F_j(X_j)$, from which the copula parameters can be estimated using likelihood methods.  We refer to \cite{czado, romeo, biauwegkamp, fermanianscaillet} for more information.

\subsubsection{Vine copulas}

As shown above, there is a plethora of flexible parametric models for two-dimensional vectors. The extensions of elliptical and Archimedean copulas to the $d$-dimensional case are much less flexible, however. Elliptical copulas impose strong constraints on the symmetry of dependence. For example, large observations cannot have stronger dependence than small observations. Archimedean copulas are exchangeable, i.e., all subsets of the variables must have the same dependence.

Pair-copula constructions were introduced by \cite{joe1996, bedfordcooke, AasCzado} to mitigate these issues. The idea is to construct a multivariate dependence structure from only bivariate copulas. Each of these bivariate building blocks describe the (conditional) dependence between a pair of variables. The bivariate copulas can be specified independently, e.g., some Gaussian and some Archimedean with different strengths of dependence, which makes these models extremely flexible.

The above construction relies on conditioning. To organize all admissible orders of conditioning, \cite{bedfordcooke} introduced a graphical model called \emph{vine}.
A vine is a set of trees, where a tree $T = (V_T,E_T)$ is a connected graph that contains no cycles, see \cite{graph}.
Regular vines (R-vines) are a nested set of $n-1$ trees $T_1, \ldots, T_{n-1}$ such that a node of the next tree ist build upon the edges of the previous tree. Furthermore the edges are only joint if they have a common node in the previous tree. If additionally the degree of each node in the first tree is at most $2$, then each tree is a path and we speak of D-vines. If each tree has a unique node of degree $n-1$, each tree is a star and we speak of C-vines. Hence D- and C-vines are special cases of R-vines. An example of a four-dimensional D-vine copula can be seen in \ref{fig:VineCop} in Section \ref{sec:num_ex}.

Every vine graph allows to decompose the joint density in a different way. To do so, each edge $e = (x, y)$ is assigned a label $(\cC_x, \cC_y | D_e)$, where $\cC_x, \cC_y \in \{1, \dots, d\}$ and $D_e \subset  \{1, \dots, d\}$. Each edge is further associated with a copula $c_e$ that captures the dependence of $(X_{\cC_x}, X_{\cC_y})$ conditional on all $\{X_k\colon k \in D_e\}$. See \cite{czado} for further details. With this notation, we get the following result.

\begin{thm}
	Let $\cX = \left(X_1,\ldots,X_d\right)$ be a sequence of random variables with continuous invertible marginal distributions $\cF = (F_1 , \ldots , F_d)$. Let $\cV$ be a R-vine tree sequence on $d$ elements and $\cB = \{ c_e ~ | ~ e \in E_i, ~ i = 1,\ldots d-1 \}$ the set of the associated bivariate copula densities. Then the joint density $f$ can be factorized into 
	\begin{align}
		f(\mathbf x) =& \left[ \prod_{i=1}^{d-1} \prod_{e \in E_i} c_{e}\left(F_{\cC_x| D_e} (x_{\cC_x } | \mathbf x_{D_e}) , F_{\cC_y | D_e} (x_{\cC_y} | \mathbf x_{D_e}); \mathbf x_{D_e} \right)
		\right]
		\cdot \left[\prod_{k=1}^d f_k(x_k) \right].
	\end{align}
\end{thm}
The conditional distributions $F_{\cC_x| D_e}, F_{\cC_y| D_e}$ in this formula can be computed recursively from the pair-copulas $c_e$ in earlier tree levels.

The formula can be simplified for a D- or C-vine. We refer to \cite{bedfordcooke, AasCzado,czado} for further reference. In this paper we use the R package \texttt{rvinecopulib} \cite{rvinecopulib}, where the latter approaches are implemented. As we don't have any knowledge about the vine structure we make use of the \texttt{vine} function, which does parameter estimation and automatic model selection using the sequential procedure proposed by \cite{dissmann}. The package also allows to simulate from the fitted model based on the inverse Rosenblatt transform algorithm \cite{dissmann,czado}.

\subsection{Dependence measures}
\label{subsec:dependence} 
One common way to measure the dependence amongst multivariate normal random variables is the Pearson correlation coefficient. This coefficient $\left|\rho_{\mathrm{BP}}\right| \in[0,1]$ is a measure of linear dependence indicating with  $\left|\rho_{\mathrm{BP}}\right| = 1$ perfect and by $\rho_{\mathrm{BP}} = 0$ no (linear) dependence. According to \cite{joe}, the coefficient $\rho_{\mathrm{BP}}$ is ``generally not the best measure of dependence'' since it may not yield $\pm$ 1 for non-Gaussian random variables with perfect dependence. 
A better choice are rank correlation coefficients, which measure monotonic dependence. Monotone association means a dependence, where if one variable increases then the other tends to decrease (or increase) and vice versa. A nice property of these measure is that they are invariant to strictly increasing transformations on the variables \cite{joe, czado}. The relation can otherwise be nonlinear.
The best known rank correlation measures are Spearman's $\rho$ and Kendall's $\tau$, which are explained in more detail in the following. 

\begin{definition}[Spearman's $\rho_s$]
Let $X_1, X_2$ be continuous random variables with continuous marginal distributions $F_1, F_2$ and let $(U_1 , U_2) := (F_1(X_1) , F_2(X_2))$. Then Spearman's rank correlation $\rho_s$ for $(X_1, X_2)$ is given by: 
\begin{align}
\rho_s := \rho_s(X_1,X_2) := \rho_{\mathrm{BP}}(U_1, U_2) = \rho_{\mathrm{BP}}(F_1(X_1), F_2(X_2)) = \mathrm{Cor}(F_1(X_1),F_2(X_2)).
\end{align}
\end{definition}
Hence, Spearman's rank correlation coefficient is defnined as Pearson's correlation coefficient of the random variables $F_1(X_1)$ and $F_2(X_2)$.
To compute Spearman's $\rho_s$ empirically from a sample $(x_{i1} , x_{i2} ), i = 1 , \dots , n$, one replaces the marginal distributions $F_1$ and $F_2$ with the respective empirical distribution functions $\widehat F_{1}$ and $\widehat F_{2}$. 
The pseudo samples $u_{ij} = \widehat F_{j}(x_{ij})$ then correspond to $r_{ij} / n$, where $r_{ij}$ is the rank of value $x_{ij}$ among all $x_{1j}, \dots, x_{nj}$. For example, $r_{ij} = 1$ if $x_{ij}$ is the smallest value in the sample.

\begin{definition}[Estimation of Spearman's $\rho_s$] 
An estimate of Spearman's $\rho_s$ based on a sample $(x_{i1} , x_{i2})$ of size $n$ with ranks $r_{ij}$ for $j = 1, 2$ is given by 
\begin{align}
\hrho_s := \hrho_s(X_1 , X_2) := \frac{ \displaystyle{\sum_{i=1}^{n}} (r_{i1} - \bar{r}_1)(r_{i2} - \bar{r}_2) }{ \sqrt{ \displaystyle{\sum_{i=1}^{n}} (r_{i1} - \bar{r}_1)^2 } \sqrt{ \displaystyle{\sum_{i=1}^{n}} (r_{i2} - \bar{r}_2)^2 } } , 
\end{align}
where $\bar{r}_1 := \displaystyle{\sum_{i=1}^{n}} r_{i1}$ and $\bar{r}_2 := \displaystyle{\sum_{i=1}^{n}} r_{i2}$ are the corresponding sample rank means. 
\end{definition}

Kendall's $\tau$ on the other hand measures dependence by comparing pairs of observations.
\begin{definition}[Kendall's $\tau$]
	Let $(X_1, X_2)$ be a random vector with joint distribution $F$ and $(X^{'}_1, X^{'}_2) \sim F$ be another vector with the same distribution, but independent of $(X_1, X_2)$. Kendall's $\tau$ is defined as
	\begin{align}
	\tau = P((X_1 - X^{'}_1)(X_2 - X^{'}_2) > 0) - P((X_1 - X^{'}_1)(X_2 - X^{'}_2) < 0).
	 \label{eq:3}
	\end{align}
\end{definition}
In the above definition, we call the pair of vectors $(X_1, X_2)$ and $(X^{'}_1, X^{'}_2)$ \emph{concordant} if the values $X_1 - X_1'$ and $X_2 - X_2'$ have the same sign. Otherwise, the pair is called \emph{discordant}. Hence, Kendall's $\tau$ is the difference of probabilites of concordance and discordance.

To calculate Kendall's $\tau$ empirically, one simply compares each observations and counts the number of concordances and ties. 

Let $N_C$ be the number of concordant pairs, $N_D$ be the number of discordant pairs, $N_1$ and $N_2$ be the numbers of ties in $X_1$ and $X_2$ respectively.
\begin{definition}[Empirical Kendall's $\tau$]
The empirical Kendall's $\tau$ is given by
\begin{align}
\htau_n^{} := \frac{N_C - N_D}{\sqrt{N_C + N_D + N_1} \times \sqrt{N_C + N_D + N_2}}.
\end{align}
\end{definition} 

As shown in \cite{czado, joe}, Kendall's $\tau$ and Spearman's $\rho_s$ are independent of the marginal distribution and only depend on the associated copula. In particular, they can be expressed as 
\begin{align}
\tau = 4 \int_{[0,1]^2} C(u_1,u_2) \,dC(u_1,u_2) -1 \quad \mathrm{and} \quad \rho_s = 12 \int_{[0,1]^2} u_1u_2 \,dC(u_1,u_2) -3.
\end{align}

%% file: numerical_examplesRound.tex
\section{Application to traffic data}
\label{sec:num_ex}
Imagine a road that is busy on one day and has little traffic on another. The driving behavior of the road users is expected to differ. For example, if we consider speed and traffic volume in terms of number of vehicles, it is reasonable to assume that these two parameters correlate. If we additionally take the time of day and weather conditions into account, we have already approached multi-dimensional dependencies that need to be considered. Since we do not only want to measure the correlation between these stochastic parameters but also want to sample or generate random values according to their associated distribution, Copulas deliver the possibility to satisfy this request. Precisely spoken we want to use these multidimensional dependencies on the driver's behavior as a stochastic input in the form of a probability distribution for the validation of autonomous systems and automated driving functions. To examine this we focus on an explicit example, described in the following.

\begin{figure}[ht!]
	\centering
	\includegraphics[scale=0.4]{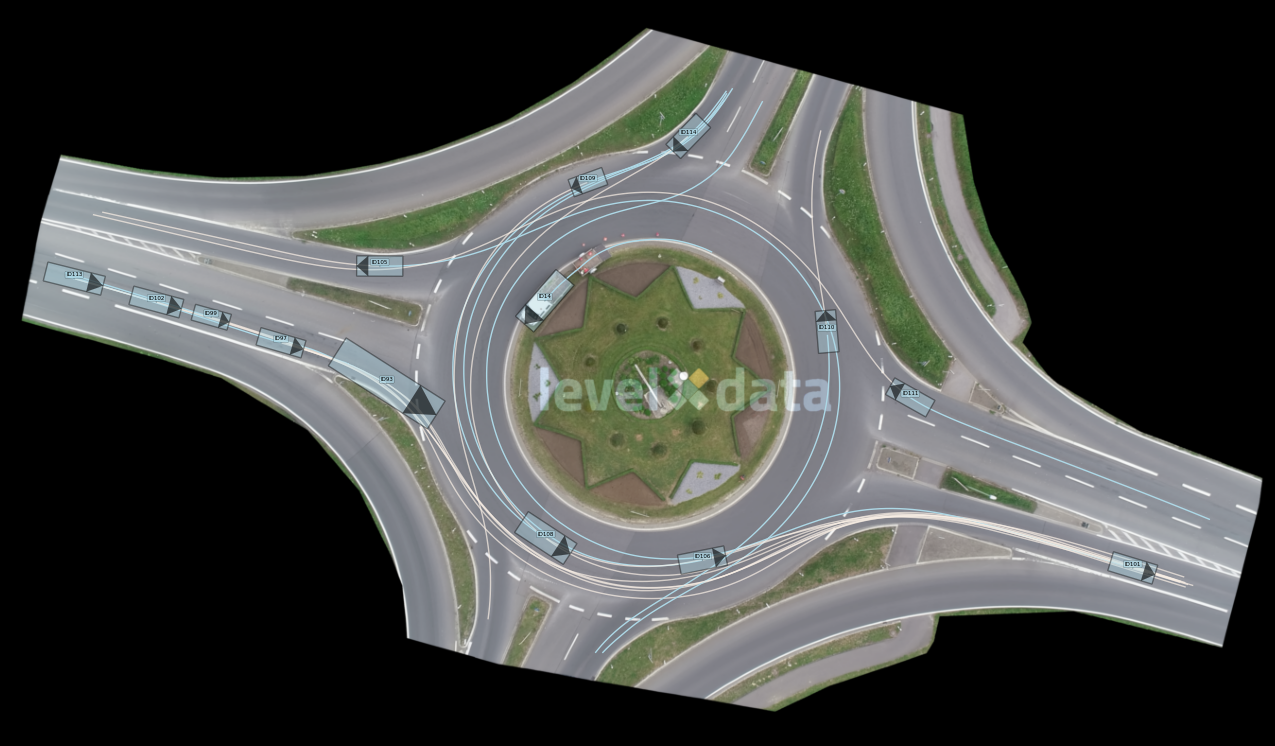}
	\caption{Recording at \textit{Neuweiler}.}
	\label{Neukoellner_Straße}
\end{figure}

The rounD dataset is a drone dataset, specifically created for behavior planning for automated vehicles, that can also be used for downstream safety validation. Through its high density of interaction between road users \cite{rounDdataset}, it is of special interest for the purpose and application of this work. Due to the recordings from the air, it avoids the risk of occlusion during recording, as well as the risk, that road users do not adopt natural behavior in traffic when being observed \cite{inDdataset}. This is one of the requirements that the dataset satisfies, to ensure that mutual influences are reflected. To create the trajectory-based dataset, which includes more than 13746 road users, traffic was recorded at 3 different locations in and around Aachen, Germany. The road user types can be divided into the following classes: cars $(11530)$, trucks $(1061)$, vans $(608)$, trailers $(257)$, buses $(53)$ and VRUs(Vulnerable Road User), that are pedestrians $(25)$, bicycles $(88)$ and motorcycles $(124)$.  
Most of the recordings were made in \textit{Neuweiler}, hence we choose this location for our application. It is a four-armed roundabout where all access roads are two-lane and the exits are single-lane roads. The fact that there is no lane marking whithin the roundabout leads to a lot of interaction. Figure \ref{Neukoellner_Straße} illustrates a recording from this location. The trajectories of the traffic participants movement up to this point are drawn in blue and the continued trajectories are drawn in white, respectively.

According to the report of the Insurers Accident Research \cite{Unfallforschung}, the most frequent accidents inside traffic circles occur due to approaching waiting vehicles too closely and disregarding the right of way when entering the traffic circle. 
\begin{table}[h!]
	\begin{center}
		\caption{Parameters}
		\label{tabPar}
		\begin{tabular}{| c | c | c |}
			\hline
			Parameter & Description & Unit\\
			\hline
			\textsf{VelCar}& velocity of cars per frame & $[\mathrm{m}/\mathrm{s}^2]$\\
			\hline
			\textsf{TrafficCar}& traffic density per car per frame & $[-]$\\ 
			\hline
			\textsf{WaitTime}& waiting time per car before entering roundabout & $[\mathrm{s}]$\\
			\hline 
			\textsf{DistCar}& minimal distance for one car to the other around him & $[\mathrm{m}]$\\
			\hline
		\end{tabular}
	\end{center}
\end{table}

Based on this evaluation, we derive the four parameters shown in Table~\ref{tabPar}. We consider the speed behavior \textsf{(VelCar)} per frame, that is already provided by the RounD data set. To determine the traffic density \textsf{(TrafficCar)}, we count the vehicles within an radius of $10$m around each car per frame. Furthermore, we consider the time of standstill \textsf{(WaitTime)} for each vehicle before entering the roundabout as well as inside the roundabout. And finally we take the minimal distance \textsf{(DistCar)} per frame, from one car to all cars in its associated surrounding. 
A total number of $132409$ samples result from the $22$ Neuweiler recordings. Each recording is taken with 25 frames per second and has a length between 19 and 23 minutes. Since we count the vehicles for the representation of the traffic density, \textsf{TrafficCar} is a discrete parameter, whereas \textsf{VelCar}, \textsf{WaitTime} and \textsf{DistCar} are continuous parameters.
\begin{table}
	\begin{center}
		\caption{Rank correlation coefficients, Kendall's $\tau$ and Spearmans $\rho$.}
		\label{tabCor}
		\begin{tabular}{| c || c | c | c |  c |}
			\hline
			\diagbox{$\tau$}{$\rho_S$} 
								& \textsf{TrafficCar} & \textsf{VelCar} & \textsf{WaitTime} & \textsf{DistCar}\\
			\hline\hline
			\textsf{TrafficCar}	& $1.0$ & $-0.45$ & $0.43$ & $-0.79$\\
			\hline
			\textsf{VelCar}		& $-0.36$ & $1.0$ & $-0.51$ & $0.49$\\ 
			\hline
			\textsf{WaitTime}	& $0.41$ & $-0.41$ & $1.0$ & $-0.39$\\
			\hline 
			\textsf{DistCar}	& $-0.64$ & $0.34$ & $-0.31$ & $1.0$\\
			\hline
		\end{tabular}
	\end{center}
\end{table}

In the lower triangular matrix of Table~\ref{tabCor}, Kendall's $\tau$ is depicted and Spearman's $\rho_s$ in the upper triangular matrix.
The values in Table~\ref{tabCor} of the rank correlations confirm, what can be intuitively surmised. If the amount of traffic participants increase (\textsf{TrafficCar}), vehicle speeds (\textsf{VelCar}) decrease, $\rho_s=-0.45$ and $\tau=-0.36$ (i.e. negative correlation), along with the distances (\textsf{DistCar}) between vehicles, $\rho_s=-0.79$ and $\tau=-0.64$ as well as waiting times (\textsf{WaitTime}) increase, $\rho_s=0.43$ and $\tau=0.41$ (i.e. positive correlation). Analogously, the speeds (\textsf{VelCar}) decrease when the waiting times (\textsf{WaitTime}) increase, $\rho_s=-0.51$ and $\tau=-0.41$ and the distances (\textsf{DistCar}) between the vehicles decrease, $\rho_s=0.49$ and $\tau=-0.34$. Furthermore, the distances (\textsf{DistCar}) decrease with increased waiting times (\textsf{WaitTime}), $\rho_s=-0.39$ and $\tau=-0.31$. 
\begin{figure}
	\centering
	\subfigure(a){
		\begin{tikzpicture}[node distance={20mm}, thick, main/.style = {draw, circle, fill=gray!30, minimum size=0.75cm}]
			\node[main] (1) {$2$};
			\node[main] (2) [right of=1] {$3$};
			\node[main] (3) [above of=2] {$1$}; 
			\node[main] (4) [right of=3] {$4$};
			\draw[-] (1) -- node[midway, above] {$2,3$} (2);
			\draw[-] (2) -- node[midway, above, sloped] {$3,1$} (3);
			\draw[-] (3) -- node[midway, above] {$1,4$} (4);
		\end{tikzpicture}} 
	\subfigure(b){
		\begin{tikzpicture}[node distance={25mm}, thick, main/.style = {draw, circle, fill=gray!30, minimum size=0.75cm}]
			\node[main] (1) {$2,3$};
			\node[main] (2) [above of=1] {$1,3$};
			\node[main] (3) [right of=2] {$1,4$}; 
			\draw[-] (1) -- node[midway, above, sloped] {$2,1\mid 3$} (2);
			\draw[-] (2) -- node[midway, above] {$1,4 \mid 3$} (3);
		\end{tikzpicture}} 
	\subfigure(c){
		\begin{tikzpicture}[node distance={30mm}, thick, main/.style = {draw, circle, fill=gray!30, minimum size=0.75cm}]
				\node[main] (1) {$3,4 \mid 1$};
				\node[main] (2) [right of=1] {$2,1 \mid 3$};
				\draw[-] (1) -- node[midway, above] {$2,4 \mid 1,3$} (2);
		\end{tikzpicture}}
	\caption{Vine tree structure}
	\label{fig:VineCop}
\end{figure}
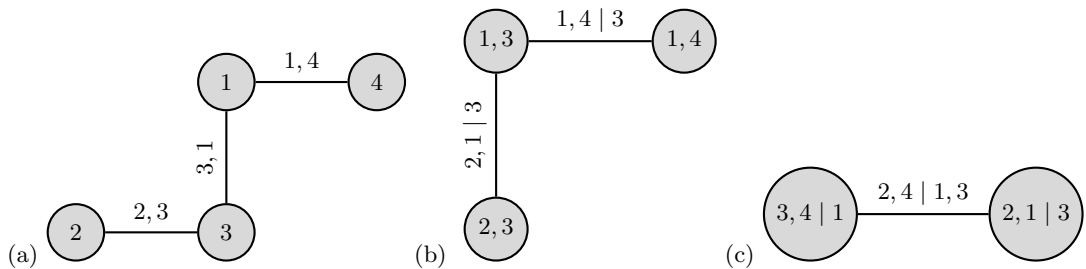
Subsequently the copula is estimated from the four parameters using the \texttt{vine} function from the \texttt{rvinecopulib} package \cite{rvinecopulib}. The resulting vine structure is shown in Fig \ref{fig:VineCop}. Therein the node $1$ represents the parameter \textsf{TrafficCar}, the node $2=$ \textsf{VelCar}, the node $3=$ \textsf{WaitTime} and node $4$ represents \textsf{DistCar}. The bivariate unconditional and conditional copula densities can be reconstructed from the tree structure such that the joint copula density $c$ can be written as:
$$
c(u_1, \ldots , u_4) = c_{23} \cdot c_{31} \cdot c_{14} \cdot c_{21\mid 3} \cdot c_{34\mid 1} \cdot c_{24\mid 13}
$$
with $u_1=\textsf{TrafficCar}$, $u_2=\textsf{VelCar}$, $u_3=\textsf{WaitTime}$ and $u_4=\textsf{DistCar}$. The unconditioned copulas $c_{23}, c_{31}$ and $c_{14}$ as well as the conditioned copulas $c_{34\mid 1}$ and $c_{24\mid 13}$ were estimated, using the local-likelihood transformation estimator (TLL). The conditioned bivariate copula $c_{21\mid 3}$ results in a (Survival) Clayton-Gumbel (BB1) family with parameters $\texttt{par}\in (0,\infty)$, $\texttt{par2} \in [1,\infty)$ and internal coding \texttt{family}: $7,17$ \cite{rvinecopulib}. 
\begin{figure}[h!]
	\centering
	\includegraphics[scale=0.56]{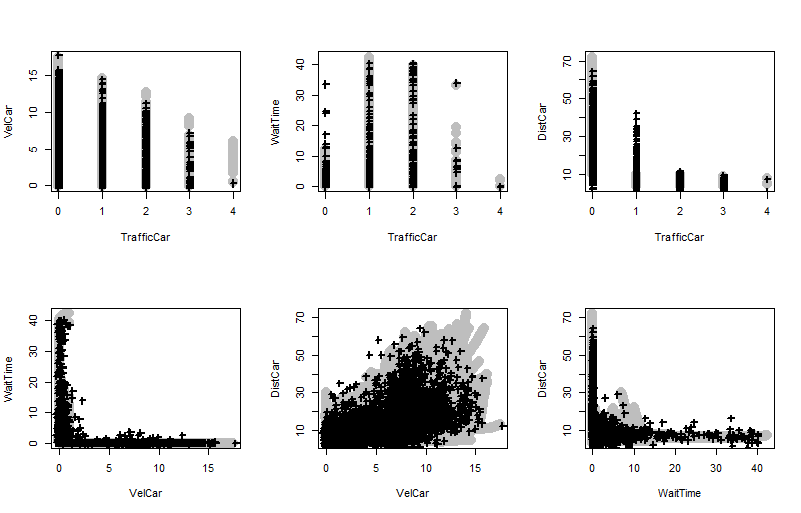}
	\caption{Measured data (grey) and associated simulated data (+) generated with the underlying copula.}
	\label{SimRealData}
\end{figure}
As mentioned before, the generation of samples similar to real data is an important requirement for reliability and risk analysis.
Figure \ref{SimRealData} shows in black $5000$ samples generated from the fitted Vine-Copula model. 
They are plotted with the according real data samples, depicted in grey, generated from the rounD dataset. Since the simulated data matches the original data sufficiently, the generated copula reflects the dependency structure of the four parameters and may therefore be used for further computation, such as the risk and reliability analysis. Further analysis is not performed in this work. In the next chapter we will give an outlook for ongoing investigation and stress tests.

%% file: conclusion.tex
\section{Conclusion}
When it comes to describing stochastic phenomena, such as the traffic behavior, the correlation between the corresponding parameters cannot be neglected. Especially, if the stochastic description of the parameters are needed for ongoing analysis, such as for the validation and verification of autonomous driving and assistance systems. In this work, copulas are used as an approach for not only measuring the dependence and correlation but also to generate and calculate the joint distribution function of multivariate stochastic variables, taking the dependence into account. 

The numerical example given in Section \ref{sec:num_ex} describes the application of the theory of copulas on a concrete observation and real-time measurements of a roundabout. Despite considering only four parameters, the speed, traffic density, waiting time (before entering the roundabout and within the roundabout) and the distance to other traffic participants, the usefulness of this approach is illustrated. Additionally, the plausibility of the negative and positive correlation of the parameters are discussed, which are captured via the use of a copula. 

It is emphasized, that more stochastic parameters can be taken into account, such as the time, since the traffic density may correlate to the time of day, e.g. rushour at the end of the working day just to give an example. Also the weather, road quality or visibility might be important parameters, depending on which questions the downstream analysis tries to address. It is worth mentioning, that more data and therefore more hours of recording is necessary to extract these parameters and to consider their interferences. The given approach may therefore be seen as an intial attempt to generate random variables or samples according to their dependencies, extracted from real traffic data, without claim of completeness of all influential parameters.

Based on the given approach, further analysis has to be performed in the future and is not considered in this work. To give an example how often an autonomous vehicle causes a crash, can be investigated with a simulated reliability analysis via e.g. Monte Carlo simulation. It is expected, that by considering the joint distribution with all correlations extracted from the real-time data, the reliability analysis is more accurate and closer to reality. But as there is no accident in the drone dataset the comparison between simulation and reality is not feasible. This highlights the need for more data and more recordings. 

This approach can also be helpful when it comes to studying the behavior of swarms of traffic participants and their interactions, similar to \cite{jrfm15070285}, especially when transfering this behavior from real-world to a virtual or simulated environment.